\documentclass[12pt]{article}

\usepackage{epsfig}

\begin{document}

\title{\bf \Large The Determination of the Metric by the Weyl and Energy-Momentum Tensors}

\author{G.S. Hall and M. Sharif \thanks{Permanent Address: Department of Mathematics,
University of the Punjab, Quaid-e-Azam Campus Lahore-54590,
Pakistan. $<$hasharif@yahoo.com$>$}
\\ Department of Mathematical Sciences,\\ University of Aberdeen,
Aberdeen AB24 3UE\\ Scotland, U.K. $<$g.hall@maths.abdn.ac.uk$>$}

\date{}

\maketitle

\begin{abstract}
This brief paper investigates the consequences for the metric
tensor of space-time when the Weyl tensor (in its conformally
invariant form) and the energy-momentum tensor is specified. It is
shown that, unless rather special conditions hold, the metric is
uniquely determined up to a constant conformal factor.
\end{abstract}

\newpage

\maketitle

\section{Introduction}

The object of this paper is to re-discuss some ideas that were
first published some time ago. In [1] and with more detail in [2]
the problem was raised as to how tightly the metric tensor of
space-time was determined by specifying the physical sources of
the gravitational field in the form of the energy-momentum tensor
together with the "free" gravitational field sources through the
Weyl tensor. In [3,4] a similar problem was raised but in a
different and more restricted form.

To establish notation let $(M,g)$ be a space-time with Lorentz
metric $g$ and Weyl tensor $C$ in its (tensor type) $(1,3)$ form
with components $C^a{}_{bcd}$. Let the curvature tensor in its
$(1,3)$ form be denoted by $\mathcal{R}$ with components
$R^a{}_{bcd}$ and the associated energy-momentum tensor in its
type $(0,2)$ form be denoted by $\mathcal{T}$ with components
$T_{ab}$. For $m\in M,~T_mM$ denotes the tangent space to $M$ at
$m$, a comma denotes a partial derivative and $\pounds$ a Lie
derivative. The metric $g$ is assumed to satisfy the Einstein
field equations. It is remarked that the Petrov type (which is a
statement involving $C$ and $g$) is actually uniquely determined
by the tensor $C$ in its $(1,3)$ form (see, e.g. [5,6]). This
result is required in what is to follow.

\section{Metric Ambiguities}

Let $g'$ be another Lorentz metric on $M$ which gives rise to the
same tensors $C$ and $\mathcal{T}$ as does $g$. Suppose also that
the Petrov type is not $O$ or $N$ over a non-empty open subset of
$M$ and that there are no non-trivial solutions for $k\in T_mM$ of
the equation $R_{abcd}k^d=0$ for each $m$ in a non-empty open
subset of $M$, where the curvature components are computed from
$g$. Then $g'=\alpha g$ for some positive constant $\alpha$ [1]
(see also [2,5,6]). It should be added that the clauses in this
statement do not prevent the result that $C$ and $\mathcal{T}$
determine the metric up to a constant factor (i.e. up to units of
measurement) being {\it generically} true (in a well defined
sense) as the results in [7] show. However, one should view this
result as being somewhat formal since it relies on the fact that
$C$ and $\mathcal{T}$ are specified in their $(1,3)$ and $(0,2)$
tensor types, respectively. One could possibly make a case for
specifying the Weyl tensor in its conformally invariant $(1,3)$
form (and which, with the above clauses on $C$, actually fixes the
conformal class of the metric - but not without them [6]) but
there seems to be no such case for the type $(0,2)$ specification
for $\mathcal{T}$.

The results in the previous paragraph involved an application of
the Bianchi identity to the curvature tensor $\mathcal{R}$ and
relied heavily on the fact that this tensor, because of the
clauses stated, is actually the same whether it is computed using
$g$ or $g'$ [1,6]. If one tries the same argument but with
$\mathcal{T}$ specified in either its $(1,1)$ (with components
$T^a{}_b$) or its $(2,0)$ (with components $T^{ab}$) form the
curvature tensors are no longer necessarily equal and
complications arise. In [3] the metrics $g$ and $g'$ {\it were
assumed conformally related} and $\mathcal{T}$ was specified in
its $(1,1)$ form. The Bianchi identity (conservation law) was then
applied to the components $T_a{}^b$ and shown to lead to a
restriction on the eigenvalues of this tensor. In fact this method
works no matters which tensor type of the energy-momentum tensor
is specified. To see this suppose the $(2,0)$ type is specified so
that $g$ and $g'$ each have type $(2,0)$ energy-momentum tensor
$T^{ab}$. If $g$ and $g'$ lead to the same Weyl tensor $C$ and if
$C$ is not of Petrov type $O$ or $N$ over any non-empty open
subset of $M$ then $g'=e^{2\sigma}g$ for $\sigma:M\rightarrow R$
[5]. If the covariant derivatives with respect to the Levi-Civita
connections arising from $g$ and $g'$ are denoted by a semi-colon
and a stroke and the associated Christoffel symbols by
$\Gamma^a_{bc}$ and $\Gamma'^a_{bc}$, respectively, then on
subtracting the equations $T^{ab}{}{}_{;b}=0$ and
$T^{ab}{}{}_{\mid b}=0$ one finds
\begin{equation}
T^{ac}P^b_{cb}+T^{cb}P^a_{cb}=0
\end{equation}
\begin{equation}
P^a_{bc}=\Gamma'^a_{bc}-\Gamma^a_{bc}=\sigma_{,b}\delta^a_c
+\sigma_{,c}\delta^a_b-\sigma_{,d}g^{da}g_{bc}
\end{equation}
On substituting (2) into (1) one finds
\begin{eqnarray}
T^{ab}\sigma_{,b}=\frac{1}{6}Tg^{ab}\sigma_{,b}\quad
(T=T^{ab}g_{ab})\nonumber\\
(\Leftrightarrow
T^{ab}\sigma_{,b}=\frac{1}{6}T'g'^{ab}\sigma_{,b}\quad
T'=T^{ab}g'_{ab}=e^{2\sigma}T)
\end{eqnarray}
Thus, whichever metric is used to compute the eigenvalues and the
trace, unless one sixth of that trace is an eigenvalue of the
associated energy-momentum tensor one is forced to conclude that
$\sigma_{,a}\equiv 0$ on $M$ and hence that $\sigma$ is constant
on $M$.

Similarly, if one assumes that the type $(1,1)$ form of the
energy-momentum tensor is the same for $g$ and $g'$ as originally
done in [3] (respectively the type $(0,2)$ form) one uses the
equations $T_a{}^b{}_{;b}=T_a{}^b{}_{\mid b}=0$ (respectively
$T_{ab}{}{}^{;b}=T_{ab}{}{}^{\mid b}=0$ and noting in this latter
case that
$T''=T_{ab}g'^{ab}=e^{-2\sigma}T_{ab}g^{ab}=e^{-2\sigma}T$) to
find
\begin{eqnarray}
T_a{}^b\sigma_b=\frac{1}{4}T\sigma_{,a}\quad
(type~(1,1)),\nonumber\\
T_{ab}\sigma^b=\frac{1}{2}T\sigma_{,a}\quad (\Leftrightarrow
T_{ab}\sigma'^b=\frac{1}{2}T''\sigma_{,a})\quad (type~(0,2))
\end{eqnarray}
where $\sigma^b=g^{ba}\sigma_{,a}$ and
$\sigma'^b=g'^{ba}\sigma_{,a}$. Again it is easily seen in each
case that the condition that the appropriate fraction of the trace
is an eigenvalue is independent of the metric used and that if
this fraction of the trace is not an eigenvalue then $\sigma$ is
constant on $M$.

\section{Specific Forms for Matter Distribution}

One can now check specific forms for the energy-momentum
distribution to see if this appropriate multiple of $T$ could be
an eigenvalue. For a perfect fluid with unit flow vector $u$,
density $\rho$ and pressure $p$ one has
\begin{equation}
T_{ab}=(p+\rho)u_au_b+pg_{ab}
\end{equation}
and so the eigenvalues are $p$ and $-\rho$ and $T=3p-\rho$. The
condition that either of these eigenvalues equals $\frac{1}{6}T$
($(2,0)$ case) are $3p+5\rho=0$ and $3p+\rho=0$, the condition
that either equals $\frac{1}{4}T$ ($(1,1)$ case) is $p+\rho=0$
(the same condition in each case) and the conditions that either
equals $\frac{1}{2}T$ ($(0,2)$ case) are $3p+\rho=0$ and
$p-\rho=0$. The dominant energy conditions [8] for the tensor (5),
assumed nowhere zero on $M$, require $\rho>0$ and $p+\rho\geq 0$
and so the condition $3p+5\rho=0$ is ruled out.

For any Einstein-Maxwell (Electrovac) field, $T=0$. But if the
Maxwell field is non-null, the energy-momentum tensor
$\mathcal{T}$ (assumed nowhere zero) is non-degenerate, being of
Segre type $\{(1,1)(11)\}$ at each point with two distinct nowhere
zero eigenvalues differing only in sign. Since in this case
$\mathcal{T}$ has no vanishing eigenvalues at any $m\in M$ a
contradiction is achieved and $\sigma$ is constant on $M$ so that
$g'$ is a constant multiple of $g$. No such contradiction,
however, is obtained if the Maxwell field is null for then
$\mathcal{T}$ has Segre type $\{(211)\}$ with all eigenvalues
zero.

If the energy-momentum content of $M$ is described by two
non-interacting perfect fluids with energy-momentum tensors of the
type (5) and pressure and density functions $p_1,~\rho_1$ and
$p_2,~\rho_2$ and fluid flow vectors $u$ and $v$, respectively,
the energy-momentum tensor is
\begin{equation}
T_{ab}=(p_1+\rho_1)u_au_b+(p_2+\rho_2)v_av_b+(p_1+p_2)g_{ab}
\end{equation}
and has Segre type $\{(1,1)11\}$ with eigenvalues
$-(\rho_1+\rho_2)-\epsilon,~p_1+p_2+\epsilon$ and $p_1+p_2$
(repeated) and where $\epsilon$ is an easily calculated positive
function of $p_1,~p_2,~\rho_1,~\rho_2$ and $u^av_a$ [9]. In this
case $T=3(p_1+p_2)-(\rho_1+\rho_2)$ and the conditions derived in
the previous section are easily written down. For example, if $T$
is prescribed in its $(1,1)$ form. they are either
$3K=-4\epsilon<0$ or $K=-4\epsilon<0$ or $K=0$ where
$K=p_1+p_2+\rho_1+\rho_2$.

Similar calculations can be performed for other combinations of
energy-momentum tensors using the algebraic results in [9] from
which it is clear that, unless rather special conditions hold, the
space-time metric is determined up to the choice of unit of
measurement. More precisely, one has the following general result.
Suppose that the type $(1,3)$ tensor $C$ is specified and is not
type $O$ or $N$ over any non-empty open subset of $M$. Then if
\\
(i) the energy-momentum tensor is specified in its $(0,2)$ form
and there are no non-trivial solutions for $k\in T_mM$ of the
equation
$R^a{}_{bcd}k^d=0$ over any non-empty open subset of $M$, or,\\
(ii) the energy-momentum tensor is specified in its $(0,2)$ form
(respectively, its $(1,1)$ or its $(2,0)$ form ) and if one half
(respectively, one quarter or one sixth) of the trace is not an
eigenvalue of it,\\
then it follows that, unless the rather special conditions
described above hold, the space-time metric is determined up to a
constant conformal factor.

\newpage

\begin{description}
\item  {\bf Acknowledgment}
\end{description}

One of us (MS) acknowledges the award of a Ministry of Science and
Technology (MOST), Pakistan postdoctoral fellowship held at the
University of Aberdeen, UK during 2002-2003.

\vspace{2cm}

{\bf \large References}

\begin{description}

\item{[1]} G.S. Hall, (19) {\it Unpublished lecture at the Fourth
Mathematical Physics Meeting Coimbra} (Portugal, 1984).

\item{[2]} G.S. Hall and A.D. Rendall, J. Math. Phys. {\bf 28},
(1987), 1837.

\item{[3]} E. Ihrig, Gen. Rel. Grav. {\bf 7}, (1976), 313.

\item{[4]} E. Ihrig, Gen. Rel. Grav. {\bf 10}, (1979), 903.

\item{[5]} G.S. Hall, in {\it Classical General Relativity} ed.
W.B. Bonnov, J.N. Islam and M.A. MacCallum (CUP, Cambridge, 1984).

\item{[6]} G.S. Hall, in the {\it Petrov Lectures at Volga 10},
(Kazan, 1998).

\item{[7]} A.D. Rendall, J. Math. Phys. {\bf 29}, (1988), 1569.

\item{[8]} S.W. Hawking and G.F.R. Ellis, {\it The Large Scale
Structure of Space-time} (CUP, Cambridge, 1973).

\item{[9]} G.S. Hall and D. Negm, Int. J. Theor. Phys. {\bf 25},
(1986), 405.

\end{description}

\end{document}